\documentclass[prl,twocolumn,showpacs]{revtex4}
\usepackage{amssymb}
\usepackage{graphicx}
\begin{document}

\title{Using NMR to Measure Fractal Dimensions}
\author{D. Candela}
\author{Po-zen Wong}

\affiliation{Physics Department, University of Massachusetts, Amherst MA
01003}
\begin{abstract}
\end{abstract}
\date{19 April 2002 v3}

\maketitle

\begin{figure}
\includegraphics[scale=0.8]{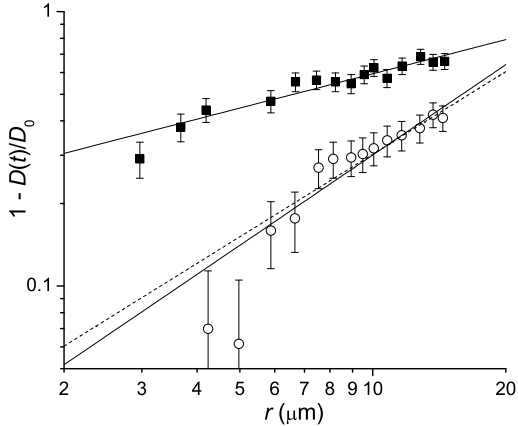}
\caption{\label{D0D}Time-dependent diffusion data for water-saturated samples of Indiana limestone (squares) and unconsolidated 15~$\mu$m dia. polystyrene beads (circles).
	The lines show least-squares fits to Eq. \ref{Dfrac2}.
	For the dashed line $D_s$ was fixed at 2.
	For the solid lines $D_s$ was allowed to vary, yielding $D_s =  2.58 \pm 0.14$ for Indiana limestone and $D_s = 2.2 \pm 0.4$ for polystyrene beads.
	To accurately determine these power laws it was necessary to separately measure $D_0$ on a bulk water sample in the same apparatus, at the same temperature.
}
\end{figure}

	In a recent Letter \cite{stallmach02}, Stallmach et al. reported pulsed
field gradient (PFG) NMR measurements of the time-dependent diffusion constant $D(t)$ in packings of water saturated sands.
	According to theory \cite{mitra93}, $D(t)$ decreases from the bulk water value $D_0$ with increasing observation time $t$ due to restrictions imposed by the pore surface:
 	\begin{equation}\label{Dsmooth}
\frac{D(t)}{D_{0}}=1-\frac{4S}{9\sqrt{\pi }V}\sqrt{D_{0}t}+\text{higher
order terms.}\end{equation}
Here $S/V$ is the surface-to-volume ratio of the pore space and $r = \sqrt{D_0 t}$ is the diffusion length.
	Stallmach et al. studied samples with different grain diameters $d_{g}$ and found that $S/V\varpropto d_{g}^{-0.7}$, which they interpreted with a fractal picture.
	If $d_{g}$ is identified as the upper cut-off scale $L$ for a fractal surface of dimension $D_s$, one expects $S/V\varpropto d_{g}^{D_{S}-3}$.
	In this Comment, we argue that the analysis of Ref. \onlinecite{stallmach02} is flawed and we propose an alternative.

	The key point is that Eq. \ref{Dsmooth} was derived for a \emph{nonfractal} surface such that $S/V$ is constant as the length scale $r$ varies.
	In Ref. \onlinecite{stallmach02}, Eq. \ref{Dsmooth} is used for $2 \ \mu\text{m} < r < 10 \ \mu\text{m}$ while $d_g$ is in the range 100-1000 $\mu$m.
	For the analysis to be valid, the surface would have to be smooth below 10 $\mu$m and abruptly turn into a fractal above this scale.
	This is implausible as sands and rocks are known to have fractal surfaces below 1 $\mu$m \cite{wong88a}.
	For example, BET measurements on rocks (for which the measurement scale is $r \approx 0.4$ nm) yield $S/V$ values one to two orders of magnitude greater than PFG NMR results \cite{hurlimann94}.

	To show how Eq. \ref{Dsmooth} is modified for fractal surfaces, we note that the term $4Sr/9\sqrt{\pi}V$ arises because molecules within a layer of volume $V_B \approx Sr$ can on average reach the pore surface within time $t$.
	Following Ref. \onlinecite{wong88} and allowing for $d_{g}>L$, it is easy to see that $V_{B}\varpropto (d_{g}/L)^{2}(L/r)^{D_s}r^{3}/(3-D_s)$.
	Hence Eq. \ref{Dsmooth} becomes
\begin{equation}\label{Dfrac}
\frac{D(t)}{D_{0}}=1-\frac{A}{3-D_s} \left( \frac{L}{d_g}\right)^{D_s -2}
\left( \frac{d_g}{r}\right)^{D_s} \left( \frac{r}{d_g} \right)^3+\text{...}
\end{equation}
where $A$ is a constant.
	This expression makes it clear that the analysis of Ref. \onlinecite{stallmach02} requires $L \varpropto d_g$, but also $r$ to be independent of $d_{g}$. 
	For a given sample with fixed $d_{g}$ and $L$, the limiting form of $D(t)$ at short times is
\begin{equation}\label{Dfrac2}
1 - D(t)/D_0 \varpropto r^{3-D_s} 
\varpropto t^{(3-D_{s})/2}.
\end{equation}

	To illustrate this method, we show some preliminary PFG NMR data on Indiana limestone and a packing of plastic beads; experimental details are given elsewhere \cite{ding96}.
	Figure \ref{D0D} shows a log-log plot of $1-D(t)/D_0$ vs. $r$.
	For a fractal surface, the data should fall on a straight line with slope $3-D_s$.
	As seen in previous studies, the data for smooth plastic beads are consistent with $D_s=2$.
	However, the limestone data fall on a line with distinctly smaller slope, giving  $D_s=2.58 \pm 0.14$.
	Optical micrographs of this limestone reveal that the pore surface has a wide range of features on scales above 1 $\mu$m.
	Although these data span less than a decade of length scales, the difference in slope between limestone and plastic beads is unmistakable.
	If the grain surfaces studied in Ref. \onlinecite{stallmach02} were truly fractal, they would show $D(t)$ time dependence similar to the limestone data in Fig. \ref{D0D}.

\bibliography{fracom}
\end{document}